\documentclass[sigconf, screen]{acmart}

\usepackage{array}
\newcolumntype{L}[1]{>{\raggedright\arraybackslash}p{#1}}

\AtBeginDocument{%
  }

\copyrightyear{2025}
\acmYear{2025}
\setcopyright{rightsretained}
\acmConference[AutomotiveUI Adjunct '25]{17th International Conference on Automotive User Interfaces and Interactive Vehicular Applications}{September 21--25, 2025}{Brisbane, QLD, Australia}
\acmBooktitle{17th International Conference on Automotive User Interfaces and Interactive Vehicular Applications (AutomotiveUI Adjunct '25), September 21--25, 2025, Brisbane, QLD, Australia}
\acmDOI{10.1145/3744335.3758494}
\acmISBN{979-8-4007-2014-7/2025/09}

\begin{document}


\title[Adaptive External Communication for Autonomous Vehicles]{Towards Adaptive External Communication in Autonomous Vehicles: A Conceptual Design Framework}

\author{Tram Thi Minh Tran}
\email{tram.tran@sydney.edu.au}
\orcid{0000-0002-4958-2465}
\affiliation{Design Lab, Sydney School of Architecture, Design and Planning,
  \institution{The University of Sydney}
  \city{Sydney}
  \state{NSW}
  \country{Australia}
}

\author{Judy Kay}
\email{judy.kay@sydney.edu.au}
\orcid{0000-0001-6728-2768}
\affiliation{
  \institution{The University of Sydney}
  \city{Sydney}
  \state{NSW}
  \country{Australia}
}

\author{Stewart Worrall}
\email{stewart.worrall@sydney.edu.au}
\orcid{0000-0001-7940-4742}
\affiliation{The Australian Centre for Robotics,
  \institution{The University of Sydney}
  \city{Sydney}
  \state{NSW}
  \country{Australia}
}

\author{Marius Hoggenmueller}
\email{marius.hoggenmueller@sydney.edu.au}
\orcid{0000-0002-8893-5729}
\affiliation{Design Lab, Sydney School of Architecture, Design and Planning,
  \institution{The University of Sydney}
  \city{Sydney}
  \state{NSW}
  \country{Australia}
}

\author{Callum Parker}
\email{callum.parker@sydney.edu.au}
\orcid{0000-0002-2173-9213}
\affiliation{Design Lab, Sydney School of Architecture, Design and Planning,
  \institution{The University of Sydney}
  \city{Sydney}
  \state{NSW}
  \country{Australia}
}

\author{Xinyan Yu}
\email{xinyan.yu@sydney.edu.au}
\orcid{0000-0001-8299-3381}
\affiliation{Design Lab, Sydney School of Architecture, Design and Planning,
  \institution{The University of Sydney}
  \city{Sydney}
  \state{NSW}
  \country{Australia}
}

\author{Julie Stephany Berrio Perez}
\email{stephany.berrioperez@sydney.edu.au}
\orcid{0000-0003-3126-7042}
\affiliation{The Australian Centre for Robotics,
  \institution{The University of Sydney}
  \city{Sydney}
  \state{NSW}
  \country{Australia}
}

\author{Mao Shan}
\email{mao.shan@sydney.edu.au}
\orcid{0000-0002-5032-6581}
\affiliation{The Australian Centre for Robotics,
  \institution{The University of Sydney}
  \city{Sydney}
  \state{NSW}
  \country{Australia}
}

\author{Martin Tomitsch}
\email{martin.tomitsch@uts.edu.au}
\orcid{0000-0003-1998-2975}
\affiliation{Transdisciplinary School,
  \institution{University of Technology Sydney}
  \city{Sydney}
  \state{NSW}
  \country{Australia}
}

\renewcommand{\shortauthors}{Tran et al.}

\begin{abstract}

External Human–Machine Interfaces (eHMIs) are key to facilitating interaction between autonomous vehicles and external road actors, yet most remain reactive and do not account for scalability and inclusivity. This paper introduces a conceptual design framework for adaptive eHMIs—interfaces that dynamically adjust communication as road actors vary and context shifts. Using the cyber-physical system as a structuring lens, the framework comprises three layers: Input (what the system detects), Processing (how the system decides), and Output (how the system communicates). Developed through theory-led abstraction and expert discussion, the framework helps researchers and designers think systematically about adaptive eHMIs and provides a structured tool to design, analyse, and assess adaptive communication strategies. We show how such systems may resolve longstanding limitations in eHMI research while raising new ethical and technical considerations.

\end{abstract}



\maketitle


\section{Introduction}

External Human–Machine Interfaces (eHMIs) are increasingly proposed as a way for autonomous vehicles (AVs) to communicate with external road users, particularly pedestrians~\cite{dey2020taming}. These systems often take the form of visual or auditory signals intended to convey the AV’s awareness, intent, or readiness to yield. However, there is growing recognition that the need for such signals is highly situational. \citet{moore2019autonomous} caution against assuming that explicit signals will always be necessary, stating that \textit{`there may be situations with ambiguous crossing priority where pedestrians strongly prefer an explicit indication of the AV’s awareness or intent'} but that signal use should be targeted to such cases. \citet{sorokin2019change} similarly argue that AV behaviour \textit{`depends on the traffic situation and only makes sense within this context.'} \citet{colley2020designspace} support this view and explicitly call for \textit{`situation-aware'} communication strategies.

Recent work also highlights the need to account for diverse user groups. \citet{wirtz2024adaptive} argue that \textit{`inclusion is a form of adaptation,'} where systems must adjust to the needs of different users. \citet{dey2024multi} highlight that relying solely on visual signals is problematic, particularly for users who are visually impaired, distracted, or facing away from the vehicle. In line with ability-based design~\cite{wobbrock2018ability}, we advocate for systems that adapt to users' abilities, rather than treating disability as a limitation to be worked around. This perspective supports the development of eHMIs that are responsive to a range of perceptual, cognitive, and situational needs, rather than optimised for an `average' user. 

Taken together, a shared direction is emerging: eHMI communication can be designed to adjust to different road users and the contexts in which it occurs. In Human-Computer Interaction (HCI), adaptive interfaces are commonly defined as systems that dynamically modify interface elements, such as layout, content, or interaction mechanisms, based on inferred user characteristics or behaviour~\cite{Mitchell1989dynamic, todi2021adapting}. Applied to the eHMI context, this means communication that adapts its form, timing, or modality based on real-time interpretation of users and contexts. In this paper, we distinguish between reactive and adaptive eHMI:

\begin{itemize}
    \item \textbf{Reactive}: Responds to a detected state with a fixed/predefined action, regardless of user or contextual variation.\\
    \textit{Example:} A fixed `You can cross' message whenever crossing intent is detected.
    
    \item \textbf{Adaptive}: The system responds and \textit{varies its communication} depending on additional, dynamically interpreted factors (e.g., hesitation, user profile, environmental conditions, presence of other road users).\\
    \textit{Example:} `You can cross' vs.\ `Take your time, you can cross': varies based on detected mobility needs.
\end{itemize}

Adaptive communication has the potential to scale across diverse traffic scenarios (e.g., multiple pedestrians, mixed traffic) and address inclusivity (e.g., supporting users with different sensory or cognitive needs). However, there remains a lack of structured guidance for how such communication can be systematically designed. 

To address this gap, we propose a framework for \textbf{actor- and context-adaptive eHMIs}. While prior eHMI research has primarily focused on pedestrians, we adopt the broader term road \textit{actors} to reflect the diversity of beings and agents AVs may encounter, including pedestrians~\cite{dey2017pedestrian}, cyclists~\cite{al2023pimp}, mobility aid users~\cite{asha2021co}, and animals~\cite{bendel2018towards}. Using the cyber-physical system (CPS) model~\cite{osswald2014prototyping, gardner2024scaling} as a structuring lens, our framework comprises three layers: \textit{Input} (what the system detects), \textit{Processing} (how the system decides), and \textit{Output} (how the system communicates). We position our contribution as a conceptual framework that supports systematic thinking and provides a structured tool to design, analyse, and assess adaptive communication strategies.

\section{Conceptual Frameworks}

A conceptual framework is a broad term referring to any structured representation of how concepts relate. In HCI, specific instances of conceptual frameworks, such as conceptual models, design frameworks, and design spaces, are commonly used to structure thinking about how systems work by identifying key components, relationships, and design considerations. Rather than prescribing fixed solutions, these frameworks support reflection, ideation, and critique. In automotive research, such frameworks are increasingly used to address emerging design challenges related to automation, inclusion, and human–vehicle interaction~\cite{colley2020designspace, dey2020taming, bengler2020hmi, detjen2022emergent, batbold2024mentorable, lidande2022building}. 

For example, \citet{lidande2022building} propose a conceptual model of perceived responsibility in automated driving. Their model is intended to support the design and development of driving automation systems that promote clearer understanding of driver responsibility during interaction with the vehicle. \citet{detjen2022emergent} present a design framework to help designers consider the needs of non-average users, such as people with physical or cognitive impairments. \citet{batbold2024mentorable} introduce a conceptual framework for `mentorability,' which highlights the evolving roles, motivations, and technologies involved in learning for older adults. Finally, \citet{colley2020designspace} propose a two-part design space for eHMIs, comprising a concept space and a situation space, which together inform the design of external communication for AVs. These examples illustrate the role of conceptual frameworks in making complex or emerging design domains more tractable, offering shared language.



\section{Framework Development}

\begin{figure*}[ht!]
    \centering
    \includegraphics[width=1\linewidth]{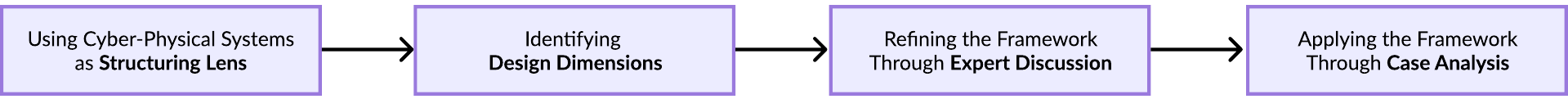}
    \caption{Overview of the framework development process.}
    \Description{}
    \label{fig:framework-development}
\end{figure*}

Conceptual frameworks can be developed through a range of approaches, including empirical insights from user or expert interviews~\cite{batbold2024mentorable, lidande2022building, detjen2022emergent, colley2020designspace}, synthesis of prior literature~\cite{batbold2024mentorable, detjen2022emergent}, and theoretical grounding (e.g., communication theory~\cite{colley2020designspace}). In this paper, we developed our framework through a four-step process (see \autoref{fig:framework-development}). 

\subsection{Using Cyber-Physical Systems as Structuring Lens}

CPS is a system where physical devices and sensors work together with computer programs to monitor the environment, make decisions, and act on those decisions~\cite{osswald2014prototyping, gardner2024scaling}. This tight link between sensing, computing, and action offers a clear way to think about autonomous systems as interconnected layers that continuously gather information, process it, and respond in real time. CPS is widely referenced in discussions of AVs and smart cities. As \citet{gardner2024scaling} explains, \textit{`autonomous systems and machines, from self-driving cars to robot vacuum cleaners, are cyber-physical systems that interconnect between hard and soft technologies, potentially multiple algorithms, and can involve forms of real-time data feedback.'}

We use CPS as a lens to structure the proposed framework into three tightly coupled layers: \textit{Input} (what the system detects), \textit{Processing} (how the system decides), and \textit{Output} (how the system communicates). This perspective makes explicit the links between sensing road actors and context, making adaptive decisions, and delivering appropriate communication through eHMIs and vehicle motion. It positions adaptive communication not as an isolated interface problem, but as a system-level capability informed by the same real-time sensing and reasoning the AV uses to control its movement.


\subsection{Identifying Design Dimensions}
\label{sec:dimensions}

To construct the framework (\autoref{fig:framework}), we synthesised insights from research on AV–pedestrian interaction and adaptive vehicle interfaces (see the research note in \autoref{sec:research-note}).

\subsubsection{Input Layer}

In HCI, adaptive systems have traditionally focused on user-adaptive interfaces—systems that tailor content, layout, or modality based on individual needs, preferences, or abilities~\cite{Mitchell1989dynamic, todi2021adapting}. In other domains, particularly safety-critical and real-time environments such as driving, contextual factors often play an equally prominent role. Therefore, we group inputs into two broad categories: actor-related and context-related.

\textbf{Actor-related inputs} are positioned along a spectrum of increasing explicitness and personalisation. Adopting this progression ensures conceptual completeness by capturing inputs ranging from general, observable cues to individualised, profile-based information:

\begin{itemize}
    \item \textit{Observed Cues} refer to directly perceivable behaviours such as gaze direction, pausing at the curb, or waving~\cite{nordhoff2025perception}. These are captured through onboard sensors without the need for inference. 
    \item \textit{Interpreted Traits} involve system-level inference about the road user's characteristics or internal state, such as estimating age, identifying the use of mobility aids, or detecting hesitation, based on observed patterns~\cite{rasouli2018autonomous}. 
    \item \textit{Actor-Initiated Signals} include explicit actions taken by road users to communicate with the AV, include active gestures~\cite{epke2021see} or requests issued through devices~\cite{tran2022designing}. 
    \item \textit{Known Profiles} may be available in cases where the AV system has access to identity-linked data, such as through a paired device or login. Known profiles could also emerge through pedestrian re-identification (re-ID) systems, where individuals are recognised across interactions~\cite{sun2024comprehensive}. 

\end{itemize}

Although our examples in \autoref{fig:framework} tend to focus on pedestrians, the framework accommodates a broader range of road actors, including non-human entities. For instance, a dog’s hesitation or a kangaroo’s presence near the roadside may prompt responsive vehicle behaviour, similar to how uncertainty is interpreted in human pedestrians. Industry explorations have proposed adapting vehicle-generated sounds to deter animals, such as high-frequency alarms triggered by GPS-based awareness of local wildlife~\cite{nissan2024roadkill}. These perspectives suggest that adaptive eHMIs could support multispecies road safety by leveraging context-aware input and tailored outputs.


For \textbf{context-related inputs}, we drew on established taxonomies that highlight the role of environmental and situational complexity in shaping interaction design~\cite{colley2020designspace, yu2024out}. These sources emphasise that context encompasses both the physical environment and the sociocultural conditions in which the interaction is interpreted. Our framework therefore includes three context-related inputs:

\begin{itemize}
    \item \textit{Environmental Conditions} encompasses physical and sensory factors such as weather, lighting, ambient noise, and infrastructure. For example, a foggy street may require stronger signals.
    \item \textit{Situational Complexity} refers to dynamic interactional factors such as traffic density, occlusion. A busy intersection, for instance, may require dynamic prioritisation or staggered eHMI cues.
    \item \textit{Sociocultural Context} captures local norms and expectations that shape how AV behaviour is interpreted, such as differences gesture use across regions.
\end{itemize}


\subsubsection{Processing Layer}

This layer builds on Feigh et al.’s Taxonomy of Adaptations for human–machine systems~\cite{feigh2012toward}, from which we adopt Adaptation Type (the `what changes' dimension describing which aspect of system behaviour or output is modified). However, its single-user focus does not fully address the complexities of AV–road user interaction. Here, external communication occurs in public, multi-actor environments. This makes it necessary to decide whether to adapt (Decision Strategy) to avoid unnecessary changes, and for whom to adapt (Personalisation Scope) to ensure messages are interpretable without disadvantaging other actors. These dimensions address adaptation challenges unique to shared road spaces that Feigh’s framework does not explicitly consider:

\begin{itemize}
    \item \textit{Decision Strategy} determines whether adaptation is warranted in the present context.
    \item \textit{Adaptation Type} specifies what aspect of the communication changes.
    \item \textit{Personalisation Scope} defines who the communication is intended for (an individual, a specific group, or all nearby actors).
\end{itemize}




\subsubsection{Output Layer}
For \textbf{eHMIs}, we drew on the design space proposed by \citet{colley2020designspace}, which offers a clear classification of \textit{Message Types}, \textit{Modalities}, and \textit{Loci} of communication. We use this structure for its simplicity and design clarity, while acknowledging the more comprehensive taxonomy introduced by \citet{dey2020taming}, which captures implementation-level variability.

In addition to eHMIs, we incorporate \textbf{kinematics} as a key output channel. Pedestrians and other road actors often interpret motion cues, such as slowing down, accelerating, or changing lateral position, as indicators of an AV’s intent~\cite{dey2017pedestrian, bengler2020hmi, rothenbucher2016ghost}. Following \citet{bengler2020hmi}, we treat motion as part of dynamic human–machine interaction (dHMI) and integrate it into the framework to ensure that physical vehicle behaviour and signal-based communication are designed in concert, rather than as separate or potentially conflicting elements.


\begin{figure*}[ht!]
    \centering
    \includegraphics[width=1\linewidth]{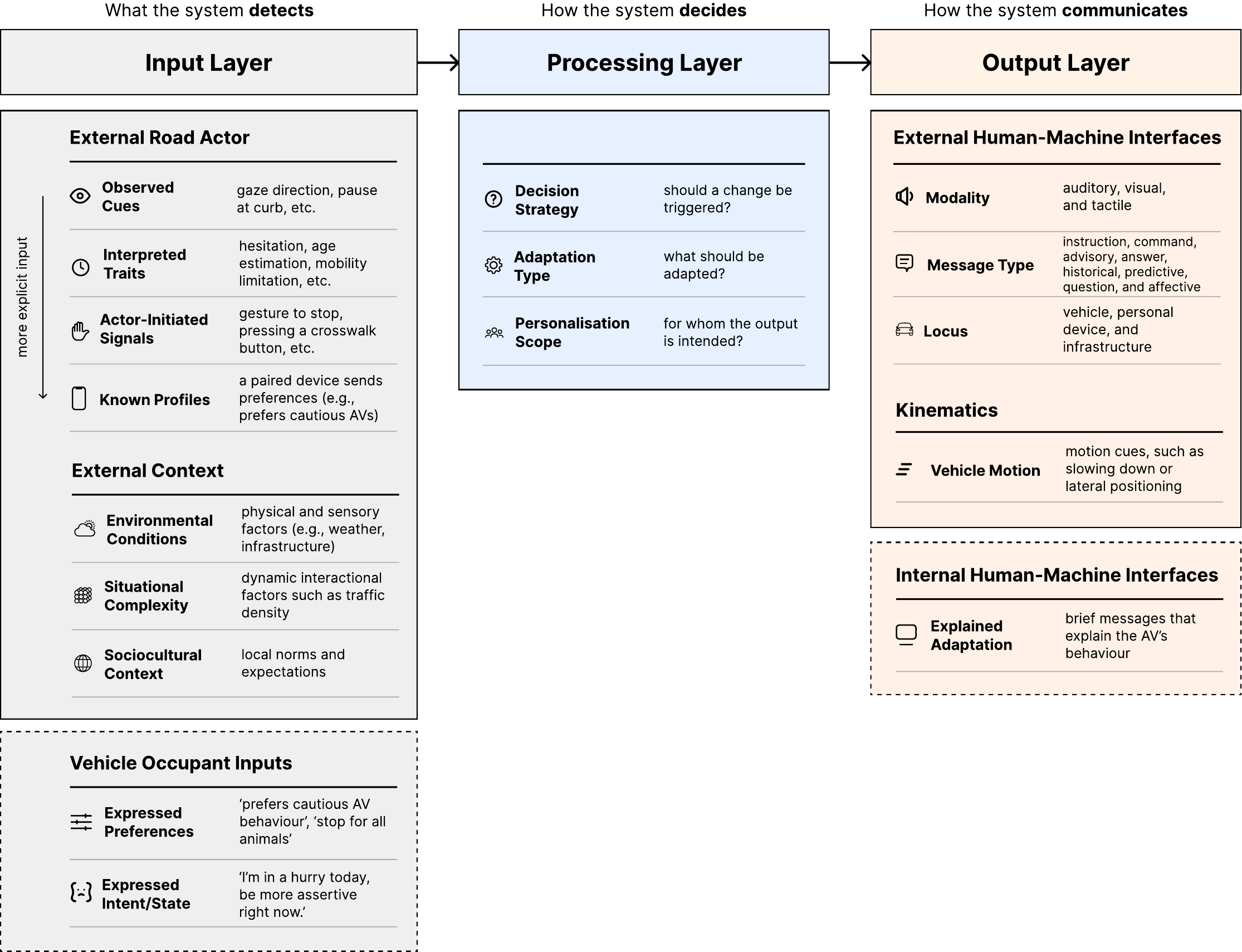}
    \caption{Overview of the proposed conceptual design framework for adaptive eHMIs. This version has incorporated feedback from the expert discussion (dotted-line areas).}
    \Description{}
    \label{fig:framework}
\end{figure*}

\subsection{Iterating the Framework: Experts Discussion}

We conducted a one-hour group discussion with seven experts in interaction design, HCI, and automotive engineering. The session focused on reviewing the framework and identifying any overlooked dimensions or practical limitations. Each expert received an A4 printout of the draft framework and its development process at the start of the session. The discussion was guided by prompts on the framework’s usefulness, completeness and clarity, and feasibility and ethics. Expert profiles and discussion questions are provided in \autoref{sec:expert-discussion}. The first author took detailed written notes during the discussion to record key points, suggestions, and examples shared by the experts. These notes were reviewed through an informal thematic process, where the insights were consolidated and mapped to specific elements of the draft framework, leading to the refinements described below.

The \textbf{Input layer} of the framework highlights a broad range of perceptual expectations that can shape eHMI adaptation. As one expert (E5) noted, it assumes ideal detection, where behaviours such as hesitation can be reliably recognised and interpreted. In practice, sensor performance may be affected by occlusion, low light, or adverse weather~\cite{combs2019automated}, and algorithms can misclassify traits like age or mobility, particularly for underrepresented groups~\cite{mehrabi2021survey}. While the framework focuses on external road users, another expert (E1) noted that internal factors such as passenger preferences may also influence adaptation. This aligns with prior work on vehicle occupant state as a design dimension~\cite{dey2020taming}, where communication may reflect the occupant’s intent or urgency. Accordingly, \textit{Vehicle Occupant Inputs} was added to \autoref{fig:framework}. Beyond the core framework, physiological or neurological data, such as stress indicators, are increasingly used in in-vehicle adaptation (e.g., \cite{meiser2022vehicle}). However, their application to external road actors remains limited due to feasibility and privacy concerns, and are therefore excluded.

Beyond detectability, whether an AV should treat a signal as relevant for adaptation is itself non-trivial. Not all inputs are equally relevant across contexts. This underscores a key challenge in the \textbf{Processing layer}: deciding not only what to sense, but when and how much it should matter in shaping communication (\textit{Decision Strategy}). Solving it may involve diverse stakeholders—pedestrians, disability advocates, urban planners—in defining what types of adaptations are acceptable or desirable. For example, adapting eHMIs based on inferred traits or behaviours surfaces tensions around privacy and consent~\cite{gardner2024scaling}. Tailoring signals for users with disabilities (e.g., low-vision pedestrians) may improve accessibility, but could also risk stereotyping. 

For the \textbf{Output layer}, while our framework focuses on adaptive communication directed outward via eHMI and vehicle motion, certain adaptations may benefit from parallel communication to passengers through internal HMI~\cite{dong2023holistic}. For example, when an AV slows or reverses to accommodate non-human road actors (e.g., wildlife)\footnote{Video `Tesla on autopilot senses three bears on highway': \url{https://www.youtube.com/shorts/9l2sf4v-13E}, last accessed 19 June 2025}, occupants may not immediately perceive the reason for this action. Minimal internal HMI cues, such as brief explanatory prompts, could enhance user trust and maintain situational transparency (E1). To account for this, we added \textit{Internal Human-Machine Interfaces} to \autoref{fig:framework}.


\subsection{Applying the Framework}


We selected two contrasting case examples to illustrate how the framework can dissect adaptive communication strategies by identifying the specific dimensions addressed, while also revealing those that may be overlooked (see \autoref{tab:case-analysis}). The first case involves sequential interactions with multiple pedestrians. The second examines mixed-traffic conditions involving both autonomous and manually driven vehicles. Together, these examples show how the framework can be applied to settings beyond simple one-to-one interactions, including those that touch on scalability to more complex, real-world traffic environments.


\textbf{Multiple Pedestrians.} ÆVITA~\cite{pennycooke2012aevita}, a pedestrian communication prototype that uses directional audio announcements to inform individuals when it is safe to cross. ÆVITA can identify multiple people using skeleton tracking and address them individually in natural language, such as \textit{`It is safe for you to walk,'} followed by \textit{`It is also safe for you to walk'} for a second person.

\textbf{Mixed Traffic.} Participants with visual impairments (VIPs) evaluated spoken eHMI messages~\cite{colley2020towards}. The preferred message, \textit{`I’m stopping, you can cross'}, was designed for vehicles with V2X (Vehicle-to-Everything) connectivity to reflect the broader traffic situation. When a manually driven vehicle was present, the message adapted to \textit{`I’m stopping'} to convey limited certainty. 

\renewcommand{\arraystretch}{1.3}
\begin{table*}[ht]
\centering
\footnotesize
\caption{Applying the framework: Two case examples.}
\label{tab:case-analysis}
\begin{tabular}{@{}L{2cm}L{3.6cm}L{3.6cm}L{3.6cm}@{}}
\toprule
\textbf{Case} & \textbf{Input Layer} & \textbf{Processing Layer} & \textbf{Output Layer} \\ \midrule

\textbf{Multiple Pedestrians (\citet{pennycooke2012aevita})} & 
Detects presence using body tracking (\textit{observed cues}) and accounts for multiple users (\textit{situational complexity}). No use of \textit{known profiles} or \textit{inferred traits}. Assumes appropriate acoustic environment and language comprehension. & 
Applies a rule-based \textit{decision strategy} to sequence announcements. Shows minor \textit{adaptation type} via phrasing (`also safe for you'). \textit{Personalisation scope} is targeted to individuals. & 
Employs directional \textit{modality} (audio), with instructional \textit{message type} and person-specific \textit{locus}. No motion-based cues; assumes vehicle is stationary. \\ \midrule

\textbf{Mixed Traffic (\citet{colley2020towards})} & 
Adapts messages based on \textit{situational complexity} (presence of mixed traffic) and \textit{user profile} (visual impairment). & 
Uses a context-aware, rule-based \textit{decision strategy} (e.g., modifies message depending on V2X input). \textit{Adaptation type} includes message content and length. \textit{Personalisation scope} is partial (broadcasted message, designed for VIPs). & 
Uses auditory \textit{modality} to support non-visual access, with blended \textit{message type} (instruction and intent). \textit{Locus} is on-vehicle. \\ \bottomrule
\end{tabular}
\end{table*}

\section{Contributions and Future Work}

As a conceptual contribution, the framework is intended not as an exhaustive or prescriptive taxonomy, but as a generative tool to support systematic thinking about adaptive eHMIs. Its primary utility lies in organising and clarifying design considerations, rather than predicting outcomes or producing empirically measurable performance improvements.  The expert discussion served as an initial check for completeness and relevance, ensuring that the framework resonates with practitioners and researchers in related fields. While this step was deliberately scoped to a concise, formative format, more formal and extensive engagement with diverse stakeholders could further refine its scope and applicability. Future iterations may reveal additional dimensions, challenge specific groupings, or propose alternative framings. 

\begin{acks}
This research was supported by the Australian Research Council (ARC) Discovery Project DP220102019, Shared-space interactions between people and autonomous vehicles. The authors thank the anonymous reviewers for their valuable comments and suggestions, which have been incorporated into the final version.
\end{acks}

\bibliographystyle{ACM-Reference-Format}
\bibliography{references}

\appendix

\section{Research Note}
\label{sec:research-note}

As noted in Section \ref{sec:dimensions} \textit{Identifying Design Dimensions}, the framework draws on research in both AV–pedestrian interaction and adaptive vehicle interfaces. This section provides a brief research note on the process.

\clearpage

\begin{itemize}
\item For the AV–pedestrian interaction literature, most of the authors, particularly the first author, who has been active in this space since 2020, draw on their familiarity with key publications and ongoing discourse in the field.
\item For the adaptive vehicle interfaces literature, a focused search was conducted on 2 June 2025 within the AutomotiveUI proceedings (via the ACM Digital Library) using the following keywords: `user modelling', `computational modelling', `adaptive HMIs', and `adaptive interfaces'. 
\end{itemize}

Due to page limits in this Work-in-Progress paper, we cite only the works that most directly informed the framework's development.

\section{Expert Discussion }
\label{sec:expert-discussion}

\subsection{Expert Profiles}

\renewcommand{\arraystretch}{1.1}
\begin{table}[ht]
\centering
\small
\caption{Profiles of experts involved in framework iteration.}
\label{tab:expert-profiles}
\begin{tabular}{@{}>{\raggedright\arraybackslash}p{1cm}%
                >{\raggedright\arraybackslash}p{6.5cm}@{}}
\toprule
\textbf{Expert} & \textbf{Area of expertise} \\ 
\midrule
E1 & Professor in Computer Science; creating infrastructures and interfaces for personalisation \\
E2 & Professor and Head of Transdisciplinary School; researches human–technology futures \\
E3 & Lecturer in Interaction Design; expertise in human–interaction design \\
E4 & PhD Candidate in Human–Robot Interaction; explores interaction with robots \\
E5 & Senior Research Fellow; studies urban interaction for vehicle automation \\
E6 & Research Fellow; specialises in perception and mapping for AVs \\
E7 & Research Fellow; works on autonomous systems, V2X, localisation and tracking \\
\bottomrule
\end{tabular}
\end{table}

\subsection{Discussion Questions}
\label{appendix:discussion-questions}

\renewcommand{\arraystretch}{1.1}
\begin{table}[ht]
\centering
\small
\caption{Questions used during expert discussion.}
\label{tab:discussion-questions}
\begin{tabular}{@{}>{\raggedright\arraybackslash}p{2cm}%
                >{\raggedright\arraybackslash}p{5.5cm}@{}}
\toprule
\textbf{Theme} & \textbf{Questions} \\
\midrule
\textbf{Usefulness} & How helpful is the framework in structuring thinking about adaptive external communication for autonomous vehicles (AVs)? \\[0.2em]

\textbf{Completeness and Clarity} & Are any key dimensions missing? Do any dimensions seem unclear, unnecessary, or confusing? Does the wording or phrasing make sense throughout the framework? \\[0.2em]

\textbf{Feasibility and Ethics} & What ethical, technical, or design feasibility concerns or opportunities does the framework raise? \\
\bottomrule
\end{tabular}
\end{table}

\section{Icon Attributions}
Icons used in this paper are from The Noun Project:

\begin{itemize}
    \item Bad Mood - Gregor Cresnar - 6450188
    \item Car - Jubayer Islam - 7882350
    \item Control - Abdullah Al Nom - 4466993
    \item Density - Megan Chown - 2892327
    \item Display - Gregor Cresnar - 7260907
    \item Eye On - Nguyen Van Sao - 2863080
    \item Gear - il Capitano - 924890
    \item Globe - Ifanicon - 4781147
    \item Hand - Amiryshakiel - 1170208
    \item Lighting - Ashwin Vgl - 6373851
    \item Message - Ahsan Damar - 7865115
    \item Motion - Gregor Cresnar - 4520057
    \item Phone - Rainbow Designs - 4084325
    \item Question - Gregor Cresnar - 670406
    \item Sound - sanjivini - 1434212
    \item Sunny - ghufronagustian - 7760760
    \item Time - Shmidt Sergey - 534804
    \item Users - Oksana Latysheva - 943303
\end{itemize}

\end{document}